\documentclass[11pt,a4paper,headsepline,footsepline,BCOR12mm,DIV13]{article}

\usepackage[utf8]{inputenc}
\usepackage[margin=1.1in]{geometry}
\usepackage{todonotes}
\usepackage[pdftex,
            pdfauthor={Gogoll et al.},
            pdftitle={Ethics in the Software Development Process: From Codes of Conduct to Ethical Deliberation},
            pdfproducer={Latex with hyperref},
            pdfcreator={pdflatex}]{hyperref}
\usepackage{subcaption}
\usepackage{amsmath}
\usepackage{amssymb}
\usepackage{amsmath,amsfonts}
\usepackage{oz}
\usepackage{algorithmic}
\usepackage{ntheorem}
\usepackage{enumitem}
\usepackage{minibox}
\usepackage{wrapfig}
\usepackage{breakcites}
\usepackage{listings}
\usepackage{xcolor}
\usepackage{graphicx}
\usepackage{multirow}

\begin{document}

\title{Ethics in the Software Development Process: From Codes of Conduct to Ethical Deliberation}
\author{
  Jan Gogoll\\
  \texttt{jan.gogoll@bidt.digital}
  \and
  Niina Zuber\\
  \texttt{niina.zuber@bidt.digital}\\
  \and
  Severin Kacianka\\
  \texttt{severin.kacianka@tum.de}\\
  \and
    Timo Greger\\
  \texttt{timo.greger@lrz.uni-muenchen.de}\\
  \and
    Alexander Pretschner\\
  \texttt{alexander.pretschner@tum.de}\\
  \and
    Julian Nida-R{\"u}melin\\
  \texttt{julian.nida-ruemelin@lrz.uni-muenchen.de}\\
}

%

\maketitle

\begin{abstract}
Software systems play an ever more important role in our lives and software
engineers and their companies find themselves in a position where they are held
responsible for ethical issues that may arise. In this paper, we try to
disentangle ethical considerations that can be performed at the level of the
software engineer from those that belong in the wider domain of business ethics.
The handling of ethical problems that fall into the responsibility of the
engineer have traditionally been addressed by the publication of Codes of Ethics
and Conduct. We argue that these Codes are barely able to provide normative
orientation in software development. The main contribution of this paper is,
thus, to analyze the normative features of Codes of Ethics in software
engineering and to explicate how their value-based approach might prevent their
usefulness from a normative perspective. Codes of Conduct cannot replace ethical
deliberation because they do not and cannot offer guidance because of their
underdetermined nature. This lack of orientation, we argue, triggers reactive
behavior such as ``cherry-picking'', ``risk of indifference'', ``ex-post orientation''
and the ``desire to rely on gut feeling''. In the light of this, we propose to
implement ethical deliberation within software development teams as a way out.

\end{abstract}

\section{Introduction}
Software systems play an ever more important role in our lives. The public
debate focuses in particular on systems that decide or support decisions about
high-stake issues that affect third parties, e.g. probation or creditworthiness \cite{o2016weapons,eubanks2018automating,noble2018algorithms}. As our reliance on software
supported decisions increases, the demand for ``ethically sound'' software
becomes more urgent. Software engineers and their companies find themselves in a
position where they are held responsible for unwanted outcomes and biases that
are rooted in the use of software or its development process (or - in case of AI
- the way the software ``learned'' (was trained) and the data that was selected
for this training or ``learning'' process). While it seems inappropriate and
short-sighted to shift responsibility entirely to developers, software companies
still feel the need to address these issues and promote ethical informed
development. This is true for two main reasons: First, companies face backlashes
from unethical software both in legal as well as in  reputational terms. Second,
companies and their employees have an intrinsic motivation to create better and
ethically sound software because it is the ``right'' thing to do. 

In this paper, we will first, briefly clarify the domain of the problem. Not
every ethical challenge a software company faces should be dealt with at the
software-developer level (or the development team). In fact, many possible
ethical issues, for instance the question if a specific software tool should be
developed at all, fall into the wider domain of business ethics. After we have
specified the possibilities and the domain of influence the software engineer
actually has regarding the implementation of ethical values, we analyze one
common approach to assist software engineers in their ethical decision making:
Codes of Ethics and Codes of Conduct.  Here we will show why CoCs are
insufficient to successfully guide SEs. We identify five shortcomings of CoCs
that make them ill equipped to provide guidance to the engineer. Finally, we
will argue that an approach built on an ethical deliberation of the software
engineer may be a way to enable SEs to build software that is ``ethically
sound''.

\section{The Responsibility of Ethical Decision Making in SE Companies}
\begin{figure}[h]
    \centering
        \includegraphics[width=\linewidth]{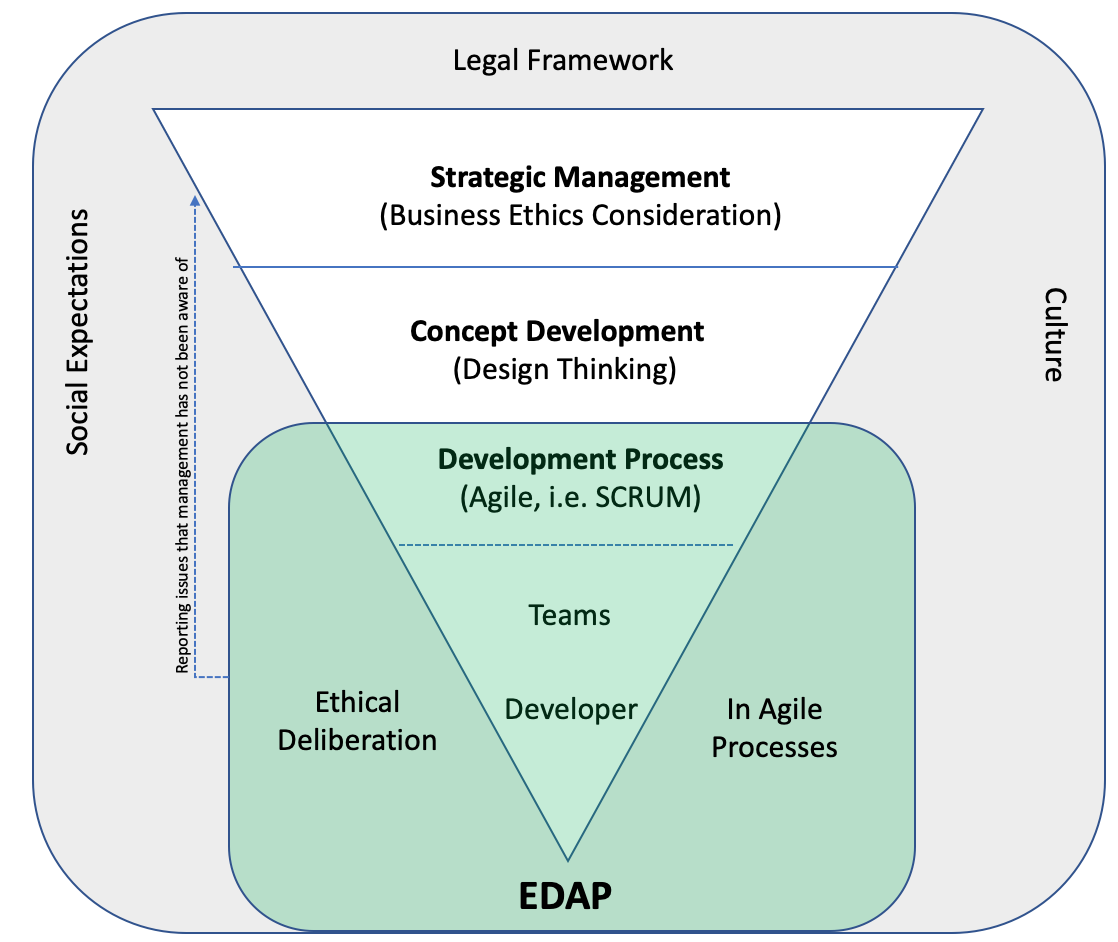}
        \caption{The domain of EDAP and the different responsibilities.}	
		\label{fig:one}
 \end{figure}

It is of crucial importance to define the domain, the scope and the limit of
ethical considerations that can be performed by software engineers and their
respective teams, before we can address the question of what ethical software
development should and can do. Many issues that seem to be the result of
software (its development and use) are actually the result of certain business
models that are deployed under certain political, legal, and cultural
conditions. Therefore, these challenges need to be addressed at the level of
business ethics rather than within the development process of software
engineers. Consider the implications for the housing and rent markets that stem
from the adoption of services such as AirBnB. This paper is not so much
concerned with these questions but with a somewhat narrower domain: After a
business decision has been made and ethical questions have (hopefully) been
pondered on the level of management, the development teams still have some
leeway in deciding how exactly to implement such a product. It is important to
note that the amount of influence of both management and development teams
changes over time. While the former has exclusive decision-making power in the
early stages (for instance the decision whether a software should be created at
all etc.), management has little control and influence in the concrete
implementation of a software product. Here the developers, armed with the
expertise on this concrete and very technical domain, realize the task or
product within the parameters set beforehand. Naturally, these parameters will
never be completely determined, thus providing the development teams with some
leeway regarding the decision of implementation.  If software engineers indeed
have the possibility to alter the concrete product in a small but maybe
ethically relevant manner, we must ask: How should they approach ethical
questions and what tools may facilitate ethical considerations? It is this
question the present paper focuses on. We developed an approach that aims  at
supporting the ethical deliberation process in software development teams and
which we call ``Ethical Deliberation in Agile Processes'' \cite{edap2020}.
Figure~\ref{fig:one} provides an overview over the different levels of ethical
decision making in a company. This graph is obviously a stark simplification of
reality, but it serves the purpose nonetheless. The fact that many decisions
have already been made before developers are assigned with the implementation of
a specific task has a great influence on the limits and the capabilities of
ethical design at the level of the software development teams.

Consider the following example which will come up again at a later point:
Technological progress enables us to build a robot that offers support to the
elderly which includes a potentially wide variety of tasks that may cover the
entire field of geriatric care. Figure~\ref{fig:one} illustrates the different layers of
ethical decision making (pertaining to manufacturing said robot). Every
organization is embedded into a web of social expectations, legal requirements
and cultural norms. In our example it is politics and a societal discourse that
establish if nursing robots are desirable at all. The eventually reached
consensus is influenced by developments such as demographic transition in
developed nations, the political goal of providing care for every senior
citizen, and the overall burden of the healthcare system. Once it has been
established that it is legal to build a specific product or artifact and a
societal consensus has more or less been achieved, the technology may be tested
(and introduced) and the management of a company can then decide to actually
manufacture such a product (for an extreme example, consider the weapon
industry). Next, a project team within the company will decide on the exact
specifications of that specific product. The nursing robot’s design can focus on
different geriatric aspects, e.g. to only assist human care workers, to be fully
autonomous in its care for an elderly human or to be deployed in specific places
only such as hospitals where the robot’s activity is subject to strict
limitations and under constant human supervision. This initial requirements
elicitation phase already addresses and decides several ethical questions. It is
then only within the narrow confines of these specifications that a developer
can and should influence the product’s design. Developers can, for example,
choose a technology that protects the user's privacy but still ensures the
achieving of established business objectives. Our care robot might use video to
interact with a patient, but it is perhaps possible to store the data in a
privacy preserving way, or even to design the robot in such a way that the data
does not need to be stored at all. Of course, if the software engineer realizes
that a major higher-level ethical issue might have been overlooked and not taken
into consideration, she has a duty to clarify and check whether the issue has
indeed been overlooked. Table~\ref{tab:one} provides an overview of different
ethical questions as well as which actors make decisions and where ethical
deliberation should be located.
\begin{table}

\begin{tabular}{ |p{3cm}|p{3cm}|p{3cm}|p{3cm}| } 
 \hline
 \textbf{Domain} & \textbf{Issue} & \textbf{Output} & \textbf{Actors} \\ 
 \hline
 1. Politics & Should the elderly be taken care of by robots? & Legal framework, societal and cultural conditions & Society \\ 
 \hline
 2. Strategy/ Business Ethics; Corporate Social Responsibility; Corporate Digital Responsibility & Should we build a robot that supports and cares for the elderly? What is the business model? & Project (approved) & Company/ Institution \\
 \hline
 3. Product Conceptualization (e.g. Design Thinking) & What capability should the robot possess? & (concrete) Requirements & Project Team, Sub division etc. \\ 
 \hline
 4. Development   Process (e.g. Scrum) & How do we implement concrete features considering the given (above) parameters? & Product & Development  Team \\ \hline

\end{tabular}
 \caption{Locating ethical questions. }
\label{tab:one}
\end{table}

Once we reach step 4, the decision to build the robot has already been made, the
business model has been chosen and concrete demands have been outlined. Any
remaining ethical questions must be dealt with by the software engineer or the
development team. Of course there are differences between companies and
corporate culture which in turn influences the degree of management’s
involvement and to what extent it fosters ethical decision making at the
development level. Yet, the developer usually has the greatest influence in
translating ethical considerations into the product, when it comes to the
concrete implementation of the task into software. If, as should be the case in
agile organizations, teams are given high-level problems, e.g., ``Find a way to
keep birds off some property'', an ethical deliberation at engineer level will
help to make explicit different options and weigh them in terms of ethical
considerations, such as: \begin{itemize}
\item A team member with a background in construction might suggest using spikes on rims and poles where birds like to sit. 
\item One with a background as a falconer might suggest getting a falcon to scare off other birds.
\item A sound engineer might suggest using high pitched noises. 
\item An environmental activist will suggest catching and relocating  the birds.

\end{itemize}

This example illustrates the possibilities and the sphere of influence of the
engineer. It cannot and should not be the case that the engineer in this example
has to decide whether it is ethically justifiable to limit the movement of birds
at all. Rather, given the constraints set at higher levels and through decisions
made earlier in the process, the engineer should focus on ethical considerations
that are in her domain and where she can assert influence.  Now that we have
established the domain in which software engineers have “ethical” influence over
an outcome, the question is then: How do we enable engineers to build software
ethically and how do we adequately consider potential ethical issues and find
solutions to these questions?

We have to acknowledge the fact that software engineers are usually not
specifically educated in ethics and have not had intensive training or other
experience in this domain. A prominent method to address the mismatch between
the lack of ethical training and the impact a product might have and therefore
the ethical attention it should receive has been the publication of Codes of
Ethics and Codes of Conduct. In the following chapter we argue that this
approach is ill equipped to achieve its intended purpose of being a useful
guideline for software engineers. Even more, we argue that ethical deliberation
cannot be delegated: neither to a machine nor by working off requirements as set
by a checklist. Consequently, we must think and weigh up ethical issues for
ourselves. Finally, we offer a proposal of how we might integrate ethical
deliberation into the very process of software development. 

\section{Codes of Conducts and Codes of Ethics}
One way of helping software engineers to detect and to deliberate ethical
questions is to establish orientation through Codes of Ethics and Codes of
Conduct (in the following the terms will be used interchangeably or simply
referred to as CoCs) in order to give ethical guidance to engineers and
management.  CoCs, for instance published by institutions such as IEEE
(Institute of Electrical and Electronics Engineers), the ACM (Association of
Computer Machinery), supranational institutions such as the EU High Level Expert
Group on AI and UNDP, or the tech industry \cite{whittlestone2019role} have a
central, (self-)regulatory function in the discourse on the development of
ethically appropriate software systems. They represent a more or less
sufficiently complete and mature surrogate of various normative positions,
values or declarations of intent, which ought to be implemented in an adequate
form in the process of software development.

The main contribution of this paper is to analyze the normative features of CoCs
in software engineering and to explicate how their value-based approach might
prevent their usefulness from a normative perspective. To this end, we identify
the most prominent  kinds of values and principles in these codes, what kind of
normative guidance they can provide, and what problems might arise from those
CoCs that uphold a plethora of abstract values. 

\subsection{Codes of Conducts, Values and Principles in AI and Software Engineering}
Codes of Ethics (CoEs) or Codes of Conduct (CoCs) are intended to provide
guidance to engineers that face ethically relevant issues and provide them with
an overview of desirable values and principles.  The ACM Code of Ethics and
Professional Conduct - for example - declares that: 

``Computing professionals’ actions change the world. To act responsibly, they
should reflect upon the wider impacts of their work, consistently supporting the
public good. The ACM Code of Ethics and Professional Conduct (``the Code'')
expresses the conscience of the profession.  The Code is designed to inspire and
guide the ethical conduct of all computing professionals, including current and
aspiring practitioners, instructors, students, influencers, and anyone who uses
computing technology in an impactful way.'' \cite{gotterbarn2018acm} Further,
the ACM Code demands that computer professionals act in accordance to their
general principles. The normative character of rules in the code suggest that
engineers should behave as indicated by them and are judged according to these
rules. Partly they are designed to be a self-commitment and partly they include
legally binding obligations which are punishable. Of course, codes of conduct
address a specific professional area and therefore remain specific in their
formulation of certain values. Nevertheless, when we consider the nominative
function of codes, the specification loses urgency, i.e. when we want to
consider the normative requirements that codes should fulfill. We find the same
normative requirements across all industries: ``They are guiding principles
designed to maintain values that inspire trust, confidence and integrity in the
discharge of public services'' \cite{secretariat2003values}. The more
precisely a range of handling can be defined, the more precise instructions can
be given: Ethics Codes of Public Relations for example are even more blurred due
unclear nature of the Public Relation domain than a specific medical ethics. 

There exists a plethora of CoCs that are addressed to software engineers in
general and developers working with artificial intelligence in particular.
Initially, CoCs were introduced by businesses as a response to growing problems
of corruption and misbehavior in business dealings. Over the years, the adoption
of CoCs has spread to many other domains, especially engineering and medicine
but also business. \cite{davis1998thinking} has argued that “a code of professional
ethics is central to advising individual engineers how to conduct themselves, to
judging their conduct, and ultimately to understanding engineering as a
profession”. CoCs would thus serve three main purposes: Firstly, they guide the
individual engineer and help to avoid misbehavior. Secondly, they serve as a
benchmark for other actors in a profession to judge potential behavior as
“unethical” and thereby contributing to the reputation of the profession as a
whole. Finally, they help to define the self-image of a profession by setting a
rulebook for what a professional actor should or should not do - this might be
particularly relevant for a comparatively young field such as software
engineering. \cite{schwartz2001nature} outlines eight metaphors that describe how
individuals may interpret CoCs: as a rulebook, a signpost, a mirror, a
magnifying glass, a shield, a smoke detector, a fire alarm, or a club (ibid.).
Questions of the CoCs’ effectiveness, however, have been raised early on.
\cite{schwartz2001nature} conducted 57 interviews in the domain of business ethics and
reported that less than half of the codes actually influence people's behavior.
\cite{kaptein2008effectiveness} find mixed results regarding the relationship
between CoCs and Corporate Social Responsibility performance of companies. More
recently and specifically targeting CoCs aimed at software engineers, \cite{mcnamara2018does} have conducted a vignette experiment to test the influence of CoCs
on developers to no avail, stating that “explicitly instructing participants to
consider the ACM code of ethics in their decision making had no observed effect
when compared with a control group” (ibid.). While the jury is still out on the
empirical effectiveness of CoCs, this paper is not overly concerned with
answering this issue directly. Rather, this article attemps to show that CoCs
conceptually fail in various ways when it comes to their main goal: Providing
ethical guidance to software engineers who find themselves in uncertain
situations.

Some research has been conducted that compares ethical codes and their values
and tries to quantify them with the goal of establishing a potential consensus.
The main focus of the current literature has been on CoCs that deal with the
development of artificial intelligence systems. \cite{zeng2018linking,fjeld2020principled,jobin2019global, hagendorff2020ethics} have analyzed what kind
of values are prominent in Codes of Ethics in the field of artificial
intelligence in order to provide an overview of the ethical principles that are
deemed important for software engineers in this particular field.  \cite{jobin2019global}, for instance, coded 84 documents within the domain of AI CoCs and
summarized eleven main principles (ordered according to the number of documents
that contain the principle, descending): Transparency, Justice/Fairness,
Non-maleficence, Responsibility, Privacy, Beneficence, Freedom/Autonomy, Trust,
Sustainability, Dignity, and Solidarity. With ``Transparency'' being mentioned
in 73 out of 84 (87\%) to ``Solidarity'' with 6 mentions (7\%), \cite{fjeld2020principled} explicitly undertook the task of analyzing CoCs to ``map a consensus''
(on the importance of principles) within the industry and the relevant
governmental and NGO players. They, too, find principles similar to Jobin et al.
They structure the content of the codes along ``themes'' which consist of values
and principles that can reasonably be subsumed under said content. They list
eight themes in total: Privacy, Accountability, Safety and Security,
Transparency and Explainability, Fairness and Non-discrimination, Human Control
of Technology, Professional Responsibility, and  Promotion of Human Values. As
mentioned above, a theme, in turn, consists of a set of principles. In the case
of ``privacy'', for instance, these principles are ``Consent, Ability to
Restrict Processing, Right to Erasure, [(Recommendation of)] Data Protection
Laws, Control over the Use of Data, Right to Rectification, Privacy by Design,
and Privacy (Other/General)`` (ibid.). \cite{hagendorff2020ethics} comes to similar
results stating that ``especially the aspects of accountability, privacy or
fairness appear all together in about 80\% of all guidelines and seem to provide
the minimum requirements for building and using an ‘ethically sound’ AI system''
(ibid.). 

\subsection{Codes of Conduct and their Normative Features}

Although there are overlaps in the listed values, such as privacy and fairness,
the recommendations for action derived from these values are quite different:
The normative concepts that are identifiable in the CoCs (henceforth: ``values'')
differ in their accentuation of content depending on the originator (NGOs, GOs,
companies, civil and professional actors) \cite{zeng2018linking}, on the addressed
product (drones, social platforms, work tracking tools, ...) as well as on the
target group (technical companies, technical professionals, civil society,
regulators, citizens). The Tech-Producer-User-Differentiation highlights that
each actor issues different CoCs targeting different interests and necessities
arising from their products or users. This means that the respective CoCs always
pertain to a certain perspective. Hence, analyzing and addressing ethical
recommendations of actions need to take these distinctions into account: origin,
product dependency, and target group. Due to this differentiation of interest
and purpose it is clear that striking differences exist in the prevalence of
values as well as in their quality. \cite{zeng2018linking} find that the average
topic frequency differs depending on the nature of the actor (government vs.
academia/NGO vs. corporations). The issue of privacy, for instance, is highly
present in government issued CoCs but (statistically) significantly lower in
academic actors and even lower in corporations. These divergences can explain
why CoCs converge on some core values, but at the same time differ tremendously
in the emphasis they put on said values as well as on the respective sub-values.
Hence, CoCs range from very abstract core-values (such as justice or human
dignity) to detailed definitions of technical approaches (e.g. data
differentiation) (see \cite{jobin2019global}). Governmental CoCs, for example,
support general and broad moral imperatives such as ``AI software and hardware
systems need to be human-centric'' \cite[p.3]{hleg2019high} without further
specification. Whereas corporations tend to favor compliance issues when taking
on privacy. 

\section{An Analytical Approach: Why Software Codes of Conducts Fail to Guide}

The majority of CoCs agree on core-values such as privacy, transparency, and
accountability. Yet, CoCs diverge as soon as this level of abstraction  must be
supplemented with application-specific details or precise definitions of
concepts. Moreover, we also encounter significant differences in the
prioritisation of values and a derivation of focal points. Given these
observations, the question is then: If CoCs are in broad agreement on core
values, why do they differ in their statements? One reasonable explanation is
that this difference is rooted in the very nature of values, namely their
underdetermination. This underdetermination is directly linked to the problem
that CoCs are barely able to provide normative orientation in software
development. This lack of orientation, in turn, triggers reactive behavior such
as ``cherry-picking'', ``risk of indifference'' and ``ex-post orientation'',
which we will discuss below. Combined, these issues result in a desire to rely
on gut feeling, so-called heuristics, that seemingly support (ethical) decision
making without much effort. Unfortunately, all those short-cuts cannot
substitute a proper ethical deliberation and thus do not allow for a
well-considered decision. In the following this criticism will be laid out in
more detail. Finally, it is argued that we must accept the challenge and
consider how we can ``master'' normativity. The final section presents a way out
of this situation. 

\subsection{The Problem of Underdetermination}

Numerous CoCs contain values that are central to the ethical handling of
software and that can hardly be reasonably disputed, such as the respect for
human dignity or the claim that technology should be developed to serve mankind
(humanistic perspective, a philosophical stance that puts emphasis on the value
and (moral) agency of human beings (see \cite{nida2018digitaler,nida2020theorie}). Although the
normativity of these values is by no means to be questioned or relativized and
these values can certainly claim normative validity, it should be obvious that a
reduction of an entire value system to these central (meta-)norms is neither
sufficiently determined in a theoretical sense nor does it lead to immediate
useful practical implications. Moreover, it is hard or impossible to deductively
derive other values from these central values. In fact, they rather take on the
role of general statements, which on their own cannot provide concrete and thus
practical guidance. A normative value system outlined within a CoC is therefore
very often underdetermined insofar as it cannot give clear instructions on what
ought to be done in any specific individual case. As a result, CoCs lack
practical applicability, because they do not offer normative orientation for
specific ethical challenges that occur on a regular basis - meaning: it fails to
achieve what it was initially created for. To make matters worse, due to the
sheer number of different values proposed in the Codes, a fitting ethical value
system to justify any possible action can be easily found since no ranking of
values can be presented with regard to the concrete and specific case at hand.

\begin{figure}
    \centering
        \includegraphics[width=\linewidth]{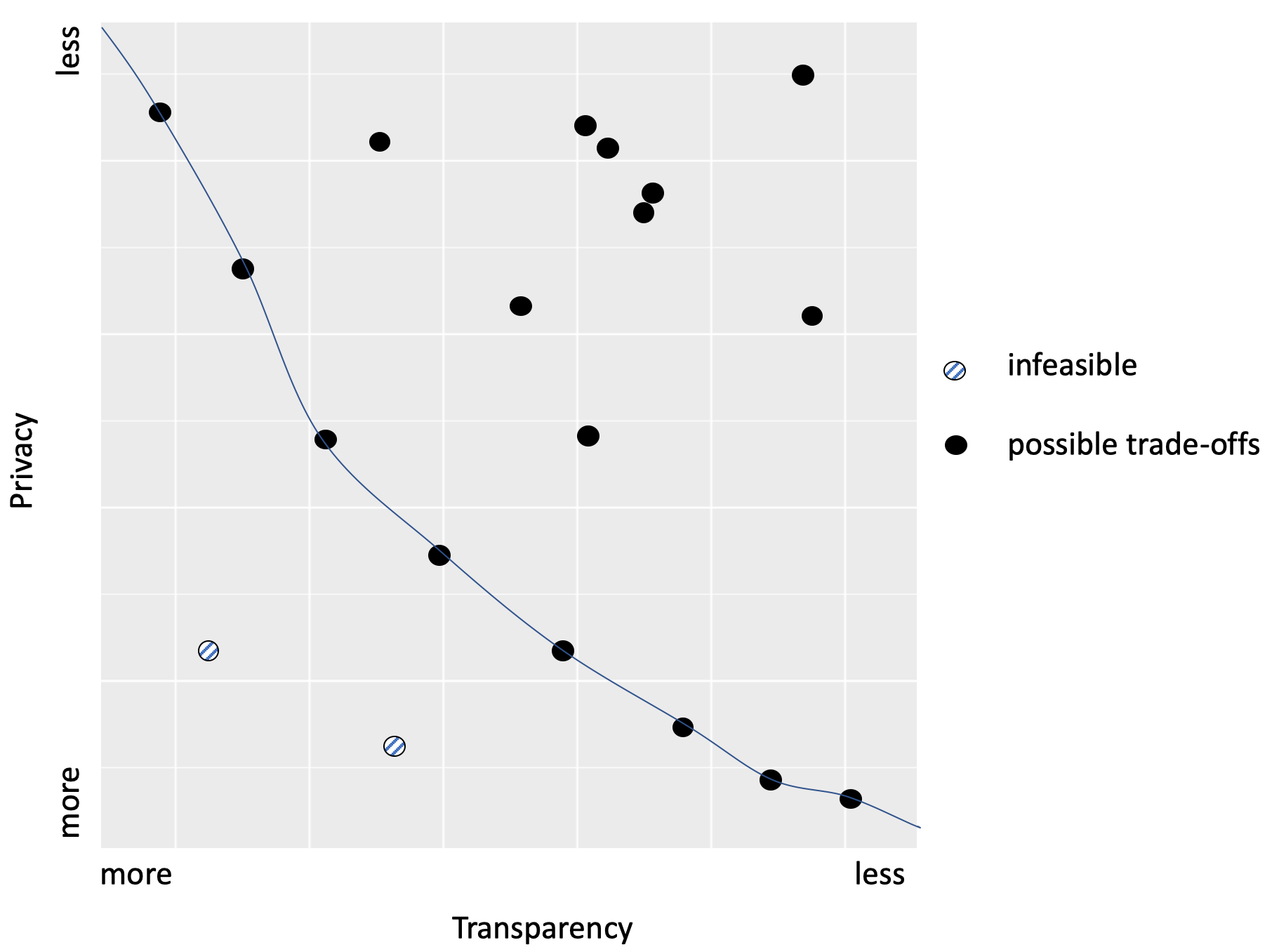}
        \caption{Trade-Offs between values and the efficiency frontier. \cite{kearns2019ethical} make a similar point about trade-offs regarding accuracy and fairness in machine learning. In their case the points would be machine learning models.}	
		\label{fig:two}
 \end{figure}

Many CoCs contain a variety of values, which are presented in a merely itemized
fashion. Without sufficient concretization, reference, contextualisation, and
explanation, software engineers are left to themselves in juggling different
values and compliance with each and everyone of them. . The nature of the values
puts them inevitably in tension with each other when applied to the reality of
software engineering (e.g. privacy vs. transparency or autonomy/freedom vs.
safety - just to name a few). More often than not, the implementation of values
ultimately requires a trade-off. Consider the example of transparency and
privacy: Both values are mentioned in the majority of the codes, yet, it is
generally infeasible to fully comply with both values simultaneously. Figure~\ref{fig:two}
shows a graphical representation of possible trade-offs between them. As long as
the product is located in the upper right corner of the graph, it is possible to
improve the situation by either increasing compliance with one value or the
other (or even both) which means moving closer towards the respective axis of
the coordinate system (here: towards the origin). Once the line is reached,
however, it becomes impossible to increase compliance with one value without
decreasing compliance with the other. This line is known as Pareto Optimality or
the efficiency frontier. While it is certainly uncontroversial that the goal of
software design should be a product that is efficiently optimizing the values we
wanted to consider, it is by no means clear or obvious which point on the line,
that is, which one of the many possible (pareto optimal) trade-offs, should be
implemented. Consider, for instance, the case of an app that enables
geo-tracking. The trade-off between privacy and transparency looks completely
different if this app is used to implement an anti-doping regime to monitor
professional athletes or as a navigation app that is used by the average citizen
to find the shortest route to her vacation destination. In the former we might
agree that professional athletes should give up more of their privacy in order
to fight illegal doping, while we are appalled by the fact that the regular user
of a navigation app is constantly tracked and monitored. 

 Yet, this is exactly the point where ethical deliberation and moral decision
 making come into play. CoCs, thus, do not offer any help to answer this
 question. In fact, they remain quiet about the very reason software engineers
 might consult them in the first place. The joint CoC from ACM and IEEE for
 instance states that ``[t]he Code as a whole is concerned with how fundamental
 ethical principles apply to a computing professional's conduct. The Code is not
 an algorithm for solving ethical problems; rather it serves as a basis for
 ethical decision-making. When thinking through a particular issue, a computing
 professional may find that multiple principles should be taken into account,
 and that different principles will have different relevance to the issue.
 Questions related to these kinds of issues can best be answered by thoughtful
 consideration of the fundamental ethical principles'' \cite{gotterbarn2018acm}. As long as we deal with win-win situations, CoCs can be applied but are
 of little use. As soon as we reach the problem of weighing legitimate ethical
 reasons and values they become rather useless. It is therefore unclear what it
 means that CoCs serve as the basis for ethical decision making when in fact the
 normative deliberation of the software engineer herself would constitute the
 footing of ethical behavior.

 \subsection{Unwanted Behavior as a Result of Underdetermination}
 \textbf{Cherry-picking Ethics}: Once Pareto optimality is achieved, any
 increase of compliance with one value must result in a decrease of compliance
 with the opposing one. It follows that in the applied case many different
 actions can be justified with recourse to various values from the same CoC
 (e.g. individual privacy vs. societal welfare). The CoC then becomes a one-stop
 shop offering an array of ethical values to choose from depending on which
 principle or value is (arbitrarily) deemed relevant in a certain situation.
 Coherent ethics, however, requiresthat ethical theory needs to cohere
 externally with our moral and general experience, beliefs, and conventions.
 Only then can ethical theory give an account for all the diversity of daily
 (normative) experiences while preserving internal coherence within an ethical
 theory e.g. using the utilitarian principle to form a consistent system of
 interrelated parts \cite{de2013ethics}. Coherent ethics cannot be realized if
 codes of conduct are understood as arbitrary accumulations of values, from
 which we can select values more or less at random. Consequently, CoCs are
 unhelpful for solving difficult decision situations as they almost always offer
 the easy way out: there will always be a value which is easy to identify or
 cheaply to apply and implement. That is why CoCs lack normative guidance. This
 is also supported by basic economic theory and experience. People usually
 choose the path of least resistance or the cheapest implementation \cite{judy2009agile}. Unfortunately, this arbitrariness and thriftiness make it virtually
 impossible to achieve a well-founded, coherent normative perspective.

 \textbf{Risk of Indifference}: Due to the shortcomings outlined above many CoCs
 seem to foster attitudes of indifference. Since they are often underdetermined
 and offer the possibility that any one particular Code of Conduct could be used
 to justify different and even contradictory actions, many Codes of Conduct
 could foster the danger of ethical indifference \cite{lillehammer2017nature}: They offer
 neither concrete nor abstract guidance and the normative function of the Code
 of Conduct is anything but guaranteed. Additionally, most Code of Conducts
 state obvious and uncontroversial values and ethical goals. In fact, their
 generic nature leaves the reader with the feeling that their gut feeling and
 practical constraints should have the final verdict when it comes to
 trade-offs. 

\textbf{Ex-post Orientation}: In addition to the problem of broadly stated
general values lacking practical orientation ("meta-values"), it is important to
understand ethics not as a restriction of an action, but as an orientation or as
an objective towards which the action should be directed, so that shared ways of
life can be supported \cite{hausman2011preference}. However, since CoCs provide values that
need to be considered but which are underdetermined due their normative
character, desired ethical values in their abstraction have little influence on
the development process. The reason for this is that values are not
process-oriented and do not include logically the means by which they can be
achieved. This very nature of values may lead to the fact that values are often
considered only afterwards, and just adapt actions, but do not align action
accordingly. This is especially true for the domain of software engineering
where every newly developed tool is very context specific and it might be harder
to bridge from abstract principles to the concrete situation compared to other
forms of engineering. It is important to stress that ethical deliberation is
more than weighing conflicting values, more than assessing consequences. We need
to think about the desirability of objectives as well as about the normative
orientation of action contexts in which technical artifacts are integrated. We
must consider normativity in the course of system development. In the course of
conceptualizing technical feasibility, we must address ethics from within since
``(e)thics on the laboratory floor is predicated on the assumption that ethical
reflection during research and development can help to reduce the eventual
societal costs of the technologies under construction and to increase their
benefits.'' \cite[p.~2]{van2013ethics}. This is precisely what we
hope to achieve with our EDAP scheme. 

\textbf{The Desire for Gut Feelings}: The underlying motivation or the desired
goal of a Code of Conduct or an ethical guideline in general could be to serve
as a heuristic. Heuristics simplify and shorten the deliberation process in
order to facilitate decisions. \cite{gigerenzer2011heuristic} describe them as
“efficient cognitive processes, conscious or unconscious, that ignore part of
the information. [...] using heuristics saves effort”. Yet, the use of
heuristics in the moral domain seems to be distinctive to their application
elsewhere, since they are based on “frequent foundation of moral judgments in
the emotions, beliefs, and response tendencies that define indignation”
\cite{sunstein2008some}. One prominent example is the connection between disgust and
moral judgement \cite{pizarro2011disgust,landy2015does}. Especially, if
novel situations and previously unseen problems arise there is little reason to
believe that a heuristic that might have been a good fit in previous cases will
also fit well into the new context. As soon as uncertainties arise because the
objectives of actions are conflicting and no unerring automatic solution can be
achieved by applying dispositions, conventions or moral rules, the resulting
lack of normative orientation must be resolved by reflection \cite{dewey2002human,mead1923scientific}. 

In sum, the underdetermination of values due to their universal character makes
it impossible  to deduce all possible specific, concrete applications of said
value. Therefore, software engineers may make an rather arbitrary and impromptu
choice when it comes to the values they want to comply with: picking whatever
value is around or - as economists would say - in the engineer’s relevant set
and which often justify actions that they want to believe to be right (this
effect is called motivated reasoning, see \cite{kunda1990case} and \cite{lodge2013rationalizing}).  And because of these two aspects - the lack of specificity as well as
the resulting cherry-picking-mentality - we encounter an attitude of
indifference \cite{spiekermann2015ethical}. Furthermore, in many cases the system is only
checked on normative issues at the end of the development process as some
technology assessment (ex-post orientation). This tendency will most likely not
lead to a change in preferences and the conceptualization of a software system
is inconsistent in terms of its normative dimension as this process disregards
normativity from the outset. Ultimately, this results in a desire for relying on
gut feeling and not putting too much thought into it  when solving all ethical
vague issues. Consequently, we encounter diffuse and unjustified normative
statements that are barely reliable. However, if we want to shape our world
responsibly, we must deliberate rationally to understand and justify what we are
doing for what cause. 

\subsection{Codes of Conduct Cannot Replace Ethical Reflection}

In this paper we argue that CoCs are insufficient to ensure or enable ethical
software development. At this point, we want to sketch out an alternative to the
reliance on abstract values in the form of CoCs and instead relocate ethical
deliberation deep within the development process. As a first step to offer
reasonable and well-founded ethical guidance, the values affirmed in the CoCs
must be made explicit and be classified with regard to their respective
context-dependent meaning as well as functional positions. Tangible conflicts
must be resolved and formed into a coherent structure that guides action
\cite{demarco1997coherence}. This process is a genuinely deliberative one which means it
cannot be reasonably expected to be successful by using heuristics or detailed
specifications that can be provided ex ante. In fact, it is not possible to
classify decision-making rules with regard to individual cases, because values
as such are context-independent and thus software engineers need assistance in
their technical development -  speaking philosophically to deliberate on issues
rather casuistically than applying ethical principles \cite{jonsen1988abuse}.
And this is exactly what we are aiming at by implementing a systematic approach
to ponder on individual cases.  For instance, the value ``privacy'' needs to be
handled differently depending on whether the context is technical or political.
While privacy issues might be addressed technically (e.g., there might exist a
possibility to store or process data while ensuring privacy) the political
discourse about the question of what level or form of data collection itself is
desirable remains unanswered. Thus, technological artefacts cannot be just
evaluated from one perspective only, but need to be assessed against the
backdrop of a multitude of categories such as authority, power relations,
technical security, technical feasibility, and societal values (see also \cite{winner1978autonomous}). Hence, an ``ethical toolbox'' in this simple form can hardly exist. The
ethical deliberation process can neither be externalized nor completely
delegated as the example of Google and their handling of the ``Right to be
Forgotten'' requests nicely illustrates \cite{corfield}. For at least some
requests to delete some piece of information from the search results it
ultimately came down  to a software engineer who flagged it as a bug and
effectively made the final decision without the legal team conducting a final
review. Therefore delegation cannot be the single solution to ethical questions
because oftentimes the ethical decision falls back on the engineer for very
practical reasons (specialists are expensive or they face an overwhelming
workload; there might even be a shortage of ethicists who possess enough domain
knowledge in software engineering). On the contrary: ethical reflection and
deliberation can and must be learned and practiced. It is a skill rather than a
checklist (see also \cite{wedgwood2014rationality}). Thus, the political goal of developing
both a general and a specifically effective CoC cannot be methodologically
separated from a practice of ethical deliberation and reflection. At least not
as long as one is not willing to give up on the notions of usability and impact.
This requires, from an ethical perspective, to raise awareness among all those
involved therefore - software developers in particular - for ethical
deliberation and its integration into the very process of the production of
software systems. Especially regarding software development, the task of
embedding ethical deliberations to agile environments with an emphasis on team
empowerment (like SCRUM) seems to be a reasonable, worthwhile, and necessary
approach (see also \cite{lopez2019systematic}). Although external ethical
expertise can be called upon to guide the common discourse, active ethical
deliberation remains irreplaceable. This is also the conclusion of the
High-Level Expert Group on Ethical AI:

\begin{quotation}
``Tensions may arise between [...] [ethical] principles, for which there is no fixed solution. In line with the EU fundamental commitment to democratic engagement, due process and open political participation, methods of accountable deliberation to deal with such tensions should be established.'' \cite[p.~13]{hleg2019high}
\end{quotation}

While \cite{mclennan2020embedded} suggestion to staff an ethicist onto each project and
every development team seems advantageous, it is hardly sensible or feasible.
Not only would this produce unjustifiable costs on software companies but the
increase in software development, a trend which will only grow in the future,
there will probably be an actual shortage of capable ethicists with a basic
understanding of software development. In order to tackle normative questions
adequately, it is therefore crucial to train software engineers in ethical
issues and to implement a systematized deliberation process. This may very well
be achieved by ethicists consulting on ethical deliberations in software
development, yet the main goal remains: the empowerment of the individual
development team regarding ethical deliberation and to implement ethics as a
skill. 

\section{Ethical Deliberation Leads to Good Normative Design}

Applied normative deliberation requires a structured, guided, and systematic
approach to the assessment of values, their trade-offs as well as their
implementation \cite{edap2020}. Reflecting ethically upon technical
artefacts to justify which features are reasonable is not a simple task since
technology is neither only a means to a given end, nor is it an end in itself.
It is also a practice that structures our social life. However, as \cite{mulvenna2017ethical}  point out, ``[w]hile most agree that ethics in design is crucial,
there is little effective guidance that enables a broader approach to help guide
and signpost people when developing or considering solutions, regardless of the
area, market, their own expertise.'' This is also due to the fact that there is
no such thing as one ethical approach that offers a strict principle or line of
thought that is applicable to all normative questions with regard to information
technological objects (e.g., this is one of the main problems with the question
of fully autonomous driving: There is no single ethical principle that fully
satisfies all normative positions and could be implemented). The field of
digital ethics already covers many different questions, ranging from computer
ethics discussing normative features of the professional ethos to machine ethics
covering questions of how to design moral machines. Moreover, in applied ethics
deontological or utilitarian principles are often used for a case evaluation.
Yet, designing objects requires identifying moral issues and to react to said
issues without limiting oneself to an in-detail argumentation that covers only a
single ethical perspective. Take our daily routines and actions: we rarely
evaluate all our options only in terms of their consequences or if they fit some
universality test as Kant suggested. Making virtues the sole basis of our
actions is also insufficient as some cases require reflection of the effects of
our actions (c.f. \cite{ross1930right,nida2002ethische,nida2020theorie}). What is more, it
is not possible to deduce all its applications logically or analytically from a
single (meta-)value. Therefore, the more software systems affect aspects of our
daily lives, the greater the demand for thoughtful engineering practice in the
form of a guided process to ensure ethically sound software. Instead of focusing
on the search for a single ethical theory that is universally applicable, we
need to introduce a practice of ethical engineering that is founded in theory.
Expertise in ethical deliberation is pivotal for identifying potential ethical
issues and addressing them properly, throughout the software development
process. However, this is not to be confounded by an understanding of ethics as
a theory or as a pure science, but rather as a type of dealing with normative
matters. 

This is what we intend to do in our framework "Ethical Deliberation in Agile
Software Processes" (EDAP) \cite{edap2020} . This approach seeks to
normatively align technical objects in a targeted manner since it enables a
goal-oriented, rational handling of values in technology design. Thus, we focus
on the identification of values, their desirability and, finally, their
integration into software systems. We try to achieve this in three steps: (1)
descriptive ethics, (2) normative ethics and (3) applied ethics. Each step must
be considered in relation to the concrete software application to be
constructed. 

\begin{enumerate} \item Descriptive ethics should facilitate access to the world
of values: which values serve as orientation? Descriptive value analysis, which
has emerged from the CoCs, is thus integrated into the deliberation process and
serves primarily as guidance. Software developers become aware of the relevant
topics within the industry in particular and society in general and to identify
their companies, societies as well as their own values regarding the object in
question.

\item Normative ethics will scrutinize the selected values, evaluate them and
subject them to an ethical analysis: Are the software features to be constructed
desirable in so far as that we would like them to be applied in all (similar)
situations (test of universality)? Do the benefits outweigh the costs given an
uncertain environment (test of consequentialism)? Which attitudes do software
developers, users or managers address explicitly and implicitly? Which
dispositions should the designed software program promote? Which desirable
attitudes are undermined (test of virtue ethics)? And finally, how do these
answers fit into the desirable life, which means the optimization of our
decision not only in regard to its consequences but also as the most
choiceworthy action in regard to the way of life we favor \cite{rawls2009theory, depaul1987two,gibbard1990wise,wedgwood2014rationality,wedgwood2017value,nida2019structural,nida2020theorie}.

\item Applied ethics, then, has to achieve even more: not only must one evaluate
the individual case in order to reach a decision, but the final normative
judgment must also account for technical possibilities and limitations - in
other words: any solution must be  technically implementable.The latter is the
interface to value sensitive design, which is an ``approach to the design of
technology that accounts for human values in a principled and comprehensive
manner throughout the design process'' \cite{friedman2002value,nissenbaum2005values,friedman2019value}. It is, therefore, a form of technically implemented
ethics and - at the same time - an imperative to the technician herself: Be
value-sensitive! For this reason we advocate the name ``normative design''.

\end{enumerate}

It is of the utmost importance that if human-machine interaction allows for
human values to be respected, technological artifacts must trace an image of the
possible normative distortions or amplifications of attitudes, rules or
behaviors that result from them: It must be made clear how artefacts structure
attitudes and actions. This calls for an ethical analysis. George Herbert Mead
emphasizes that ethics can only suggest the method of dealing rationally with
values and that these, in turn, are dependent on specific circumstances:

\begin{quotation}
 ``The only rule that an ethics can present is that an individual should
 rationally deal with all the values that are found in a specific problem. That
 does not mean that one has to spread before him all the social values when he
 approaches a problem. The problem itself defines the values. It is a specific
 problem and there are certain interests that are definitely involved; the
 individual should take into account all of those interests and then make out a
 plan of action which will rationally deal with those interests. That is the
 only method that ethics can bring to the individual.'' \cite{mead1934mind}.
\end{quotation}

To highlight the impossibility of (logical) reductiveness and thus the necessity
of rational deliberation, consider again the following scenario from the
beginning: A software and robotics engineering team is tasked to construct a
nursing roboter for the elderly. This system is supposed to take care of
individuals, i.e. help them when they fall, contact the ambulance in case of an
emergency, etc. The first step is to hermeneutically understand the scenario. In
order to create a normatively appropriate system, it is important to locate the
essential desirable values and discuss their exclusivity. To this end, one must
identify normatively desirable ``anchor points'' and consequently examine them
in regard to their relationality. Values and their interconnectedness cannot
simply be logically derived from the description of the situation. In the
nursing-robot case, we are concerned with the desirable objective of being able
to lead a self-determined life for as long as possible. Regarding technical
features, such a system may include a built-in-camera and an audio microphone to
receive voice commands to track the patient’s position in order to guarantee an
adequate reaction, e.g. calling for help in case of an emergency or assisting in
getting back up if no injury is detected. Additionally, the stored data may give
doctors the possibility to check for progression of dementia or other illnesses
or to detect potential diseases through interpretation of these images using
artificial intelligence. Much else is conceivable depending on objectives,
funding, technological progress, and potential legal issues. Normatively
speaking, the system enhances desired core values such as autonomy and
wellbeing. A nursing robot that is not only technically robust and safe, but
also constructed in accordance with the relevant normative standards still
classifies critical situations, but this is done in a respectful, humane manner.
In order to create systems that are - at least in principle - normative
adequate, we must ask which values are at risk when promoting autonomy by
implementing a camera or audio-recording system. In this specific case privacy
concerns seem dominant. Thus, we have to ensure privacy while enabling autonomy
which means, for instance, that people may not want to be filmed naked and do
not want to disclose the location of their valuables etc. Furthermore, people
may want to know who has access to the data, what kind of data is stored, why
and where. In a perfect scenario, access to the data is exclusively limited to
doctors who may use it to predict and treat illnesses (medical prevention). Even
these basic normative considerations cannot be logically derived from the
premise of promoting welfare and autonomy. We need to reason normatively to
understand and highlight the ethical issues at hand while simultaneously using
our empirical knowledge of the world. Consequently, technical solutions such as
visual or audio recording systems must be developed in such as way  that they
would not record certain scenes at all or in case, where recording is medically
necessary, recording must ensure a fair autonomy-privacy ratio by using specific
techniques such as cartooning or blurring \cite{padilla2015visual}. Other normative
issues result from using certain techniques that may meet transparency
requirements. Users want to know how recommendations are made and what reactions
they can expect. A non-technical normative deliberation is whether it is
desirable to live with an assistance system at home by oneself even if the
system considers relevant normative aspects. These kinds of ethical concerns are
not addressed within our ethical deliberation tool as we already outlined at the
beginning. We focus on software development from the engineer's perspective.
Whether or not there should be a nursing robot in the first place is a question
that falls into the domain of (business) ethics and cannot be decided on the
level of the software developer. As mentioned at the beginning, management needs
to tackle these questions, and consider the legal framework as well as other
basic norms of society. The ethical deliberation of the developer, however,
begins with the specific implementation of the technical object. For instance:
The above-mentioned need for visual observation seems to be obvious - after all,
the robot needs to ``see'' its surroundings in order to be a useful tool. Yet,
there are many ways to implement the visual capabilities of the robot and to
safeguard the privacy of individuals (cartooning, blurring etc.). At this point
the software developer usually has the freedom (and therefore the
responsibility) to deliberate on which specific implementation to use. 

At this point we would like to take it one step further than  van der \cite{van2013ethics} and highlight that not only societal costs may be reduced and
profits of the developing firm may be increased. Taking normative issues
seriously from the very beginning of a development project may lead to more
effective systems and  processes - especially in the long-run. We must
understand that technology transforms and systematizes our lives.  And that it
does so with a certain requirement for behavioral adaptation by the user or
those affected by the system in order to ensure the functionality of the system.
Hence, technology structures normatively the world we live in. This means we
need to take normativity as we encounter it in our daily lives seriously and
think about the technical artefact as being a part of it, i.e. only then we can
understand that an ethical deliberation is not only a deliberation of
trade-offs. Moreover, we can highlight how the artefact can be normatively
integrated into our daily routines. This means also to think about the
compatibility of algorithms, data sources and inputs, such as control commands,
but also front-end design, such as readability to make sure that some desired
normative features are technically well met \cite{simon2012democracy,friedman2019value}.
Normativity, thus, is not to be understood as some qualifier, but rather
as a condition that enables specific desirable practices of daily life \cite{rip2013pervasive}.

\section{A Paradigm Shift: From CoCs to Ethical Deliberation in the Software Development Process }

CoCs are difficult to use as normative guidelines for technical software
development due to their underdetermined character. They may trigger behavior
such as indifference or the cherry-picking of specific ethical values. Thus,
they are of little immediate use during the software development processes.
Since CoCs lack direct real-world applicability, ethical values may only be
chosen after the product is finished depending on which values ``fit'' (are most
compatible with the existing product). However, in order to build an ethically
sound system, it is essential to consider normative issues during the
development process. Only when integrating values from within the development
process, engineers will be able to consciously build software systems that
reflect normative values. This will foster an understanding of how technological
artefacts act as normatively structuring agents and improve the chances of
creating ethically informed software systems. In contrast to related work, we
suggest moving from a simple application of ``ethical heuristics'' to a point
where we treat ethical thinking as a skill that has to be practiced and can be
embedded deeply into the software development process. Consequently, ethical
deliberation must not be limited to ethics councils, company advisory boards or
other special committees. Rather, it needs to be practiced and shaped by the
software developers who create and intricately understand the technical system.
This approach would lead to ethical empowerment of software engineers, as part
of the empowerment trend that we see in agile software development anyway. It is
important to outline that developers should only be concerned with ethical
issues that belong to their field of competence and the area they can actually
affect. Ethical issues outside the scope of a single developer can then be
delegated to other instances of the organization. In this sense the ethical
deliberation of software developers can also lead to a positive form of whistle
blowing when an ethical issue is detected during the implementation process that
has gone unnoticed by decision makers of the company.

\bibliographystyle{apalike}
\bibliography{bibliography.bib}

\end{document}